# HgTe quantum wells for QHE metrology under soft cryomagnetic conditions: permanent magnets and liquid $^4$He temperatures

I. Yahniuk,[1] A. Kazakov,[2] B. Jouault,[3] S. S. Krishtopenko,[1,3] S. Kret,[4] G. Grabecki,[4] G. Cywiński,[1] N. N. Mikhailov,[5] S. A. Dvoretskii,[5] J. Przybytek,[1] V. I. Gavrilenko,[6] F. Teppe,[3] T. Dietl,[2,7] and W. Knap[1,3,*]

[1]*CENTERA Laboratories, Institute of High-Pressure Physics, Polish Academy of Sciences, PL-01142 Warsaw, Poland*
[2]*International Research Centre MagTop, Institute of Physics, Polish Academy of Sciences, al. Lotników 32/46, PL-02668 Warsaw, Poland.*
[3]*Laboratoire Charles Coulomb, Université de Montpellier, Centre National de la Recherche Scientifique, F-34095 Montpellier, France.*
[4]*Institute of Physics, Polish Academy of Sciences, al. Lotników 32/46, PL-02668 Warsaw, Poland.*
[5]*Institute of Semiconductor Physics, Siberian Branch, Russian Academy of Sciences, pr. Akademika Lavrent'eva 13, Novosibirsk, 630090 Russia.*
[6]*Institute for Physics of Microstructures, Russian Academy of Sciences, N. Novgorod, 603950 Russia.*
[7]*WPI-Advanced Institute for Materials Research, Tohoku University, Sendai 980-8577, Japan*

**Abstract**

HgTe quantum wells with a thickness of ~7 nm may have a graphene-like band structure and have been recently proposed to be potential candidates for quantum Hall effect (QHE) resistance standards under condition of operation in the fields above certain critical field $B_c$, above which the topological phase (with parasitic edge conduction) disappears . We present experimental studies of the magnetoresistance of different of HgTe quantum wells as a function temperature and magnetic field, determining the critical magnetic field $B_c$. We demonstrate that for QWs of specific width $B_c$ becomes low enough to grant observation of remarkably wide QHE plateaus at the filling factor $\nu = -1$ (holes) in relaxed cryomagnetic conditions: while using commercial 0.8T Neodymium permanent magnets and temperature of a few Kelvin provided by $^4$He liquid system only. Realistic band structure calculations evidence that the observed phenomena can be qualitatively understood by considering two types of the charge carries – the light and heavy holes. The formers are attributed to the Γ point of the Brillouin zone, while the latters are caused by the side maxima of the valence band. Our work clearly shows that the peculiar band structure of HgTe QWs make them a favorable platform for developing metrological devices under soft cryomagnetic conditions.

## I. INTRODUCTION

The quantum Hall effect (QHE) is arguably the first topological phenomenon that has important practical applications. In a two-dimensional electron gas (2DEG) submitted to a high perpendicular magnetic field, the Hall resistance develops steps separated by plateaus whose value is quantized and given by $\rho_{xy}^{\text{plateau}} = h/(ie^2)$, where $R_K = h/e^2$ is the von Klitzing resistance, $i$ is an integer, $h$ and $e$ are the Planck constant and the electron charge, respectively [1,2]. Thus, the QHE relates an electrical resistance to fundamental physical constants. At the same time, QHE measurements can be so precise that the resistances standards, currently maintained in various national metrology institutes, are based on the QHE and have a relative precision of a few parts per billion [3]. This precision was one of the triggers behind introducing the *quantum* International System of Units (SI), in which $h$ and $e$, and therefore $R_K$ have fixed values [4].

For these metrological applications, the choice of the material is crucial. To reach ideal quantum limit conditions, one needs the highest possible Landau level (LL) splitting (low carrier mass and high magnetic fields), the lowest possible broadening of the LLs and of the Fermi-Dirac distribution function (high mobility, low temperatures). Up to recently, only Silicon metal-oxide-semiconductor field-effect transistors (MOSFETs) and GaAs/AlGaAs 2DEG heterostructures could be used for accurate measurements of the quantum Hall resistance. Usually, GaAs devices are preferred because of their superior properties [4-7]. However, both types of devices are accurate only with stringent experimental conditions: they operate at extremely low temperatures, $T \simeq 1.5$ K or below, and under high magnetic fields, $B \simeq 10$ T that can be provided only by superconducting magnets.

It has recently been demonstrated that graphene can operate as a quantum Hall resistance standard (QHRS) with quantization accuracy of a few parts per billion [8-10]. Furthermore, graphene can even work under less demanding experimental conditions than GaAs or Si semiconductor-based standards [9-10]. This results mainly from the significant LL energy separation which can be reached in graphene, thanks to the peculiarities of its energy band structure, which mimics the energy-momentum relation of massless Dirac fermions and has a high Fermi velocity [11].

However, graphene is not the only material in which large LL energy gaps can be obtained at low fields. This distinct behavior is an inherent property of all systems hosting massless two-dimensional (2D) Dirac fermions. HgTe/HgCdTe quantum wells (QWs) are structures in which the 2D confinement can be tailored to obtain a Dirac-like band dispersion as in graphene [12-14]. Indeed, with the decrease of the QW width, one may change the inverted band structure of a HgTe QW to a normal one. For rectangular HgTe/Cd$_{0.7}$Hg$_{0.3}$Te QWs grown on CdTe buffer, the band dispersion of massless Dirac fermions is realized at the critical thickness $d_c \sim 6.3$ nm [13-18]. For example, the computed band structure for two rectangular HgTe quantum well structures of thickness $d = 7.0$ nm, and $d_c = 6.3$ nm are shown in Figs. 1(a,b). The peculiar behavior of two Landau levels in HgTe QWs with the inverted band sequence, referred to as "zero-mode" Landau levels, leads to their crossing at a certain critical magnetic field $B_c$ [see Fig.1(c)]. According to Figs. 1(a,c), below (above) $B_c$, the QW hosts a topological (normal) bandgap. In the case of the Dirac-like energy dispersion, the critical field is zero, $B_c = 0$, as shown in Figs. 1(b,d). Details of the calculations and parameters are given in Ref. 19.

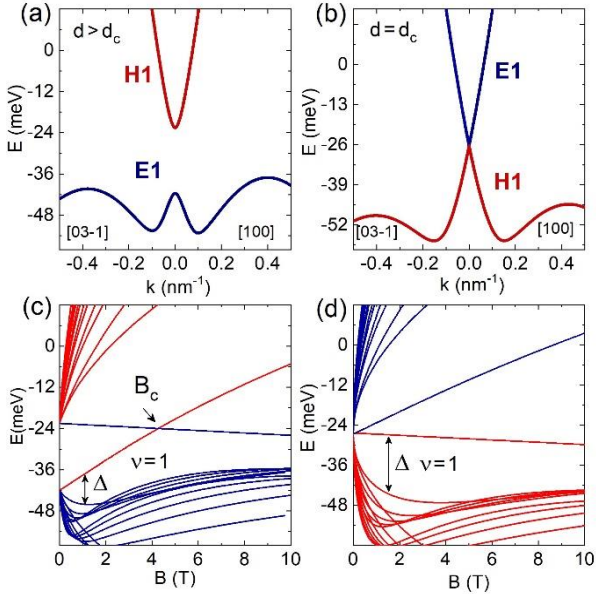

**Fig. 1.** Calculated band structure and Landau levels for rectangular HgTe quantum wells of (a,c) $d = 7.0$ nm and (b,d) $d_c = 6.3$ nm thickness, respectively. The red curve (H1) corresponds to a heavy hole-like subband, and the blue curve (E1) corresponds to an electron-like subband. Symbol $B_c$ is a critical magnetic field in which zero-mode LLs cross. The magnitude of $\Delta$ determines the energy gap between the two uppermost LLs originating from the valence band.

Thanks to the recent progress of molecular beam epitaxy (MBE) growth, HgCdTe-based QW structures with high carrier mobility $\mu > 10^5$ cm$^2$/Vs could be obtained. These structures allowed for the observation of the QHE with the filling factor $\nu = 1$ in relatively low magnetic fields ($B < 1$ T) [14-16]. In Refs. 17 and 18, the QHE in HgTe QWs was observed even up to liquid nitrogen temperatures. These observations allow considering HgTe/HgCdTe QWs as promising candidates for QHRSs. It has been predicted in Ref. 20 that the QHE standards might be realized in relaxed cryomagnetic conditions: without any superconducting magnets and without any sub-Kelvin refrigerators required hitherto to observe the quantized Hall resistance in weak magnetic fields [21-23].

In this work, we report on experimental studies in slightly inverted HgTe QW structures in view of their possible application for QHRSs. We demonstrate that for holes, one can reach the QHE states at $\nu = -1$ in relatively favorable conditions: magnetic fields below 1 T and temperatures up to 10 K. We also show essential differences in the QHE plateaus widths between holes and electrons, and the temperature-induced transition from the inverted to normal band ordering. Finally, we present the first demonstrator of the QHE resistance standard using Nd permanent magnets providing magnetic field $B = 0.82$ T, placed in $^4$He cryogenic system that can be very convenient for educational purposes as well as for some less demanding metrological applications.

We also discuss the necessary steps to reach with HgTe QWs the ultimate/high metrological precision keeping relaxed cryomagnetic conditions. Our work clearly shows that the peculiar band structure properties of HgTe QWs with massless Dirac fermions make them an ideal platform for developing metrological devices.

## II. EXPERIMENTAL

Our experiments were carried out in HgTe QW devices with top gate electrodes. The investigated (HgTe/Hg$_{0.38}$Cd$_{0.62}$Te) structures were grown on GaAs (013) substrate by MBE and had QW thicknesses 7.5 nm (device D1) and 7.1 nm (device D2) [24]. First, Hall bars with dimensions $L \times W = 5 \times 3$ μm$^2$ (device D1) and $L \times W = 40 \times 20$ μm$^2$ (device D2) were patterned by electron beam lithography. Then, a Si(ON) dielectric layer of thickness 140 nm was deposited onto the sample surface. Finally, a metallic gate completed the structure. The processing protocol was previously described in detail [25].

Magnetotransport experiments have been performed in a custom-made $^4$He cryogenic system in the temperature range of 1.8 – 20 K and magnetic fields up to 5 T provided by a superconducting coil. We have used SR830 lock-in amplifiers with low modulation frequency (~13 Hz) to determine the longitudinal and the Hall components of the resistance tensor, $R_{xx}$ and $R_{xy}$, respectively. The current flowing through the devices has been limited to the range 10 nA – 1.0 μA.

In the experiments with permanent magnets, the sample was mounted in a Teflon spacer of thickness 2.5 mm inserted between two pairs of permanent Nd magnets (two circular cylinders with diameter $D \approx 22$ mm and height $h \approx 10$ mm) that providing an axial magnetic field $B = 0.82$ T at their center. Using finite element modeling, we estimated the field homogeneity to be around 2% in a radius of 5 mm.

## III. RESULTS

Four probe measurements of $R_{xx}$ and $R_{xy}$ have been carried out for various gate voltages $V_g$ at $B = 0.2$ T to determine carrier densities and mobilities. The Hall concentration was experimentally determined as $n_H = 1/eR_H$, where $R_H$ is the Hall coefficient. As in the case of field-effect transistors, 2D carrier density is also given by $n = \alpha(V_g - V_0)$, where $V_0$ is a gate offset, and $\alpha$ depends on both geometric and quantum capacitances per surface unit. As the valence band side maxima trap holes, α is higher on the electron side ($\alpha_n$) than on the hole side ($\alpha_p$). For device D1, $\alpha_n \approx 1.4 \times 10^{11}$ cm$^{-2}$ V$^{-1}$ (and $\alpha_p \approx 1.2 \times 10^{10}$ cm$^{-2}$V$^{-1}$), whereas for device D2 $\alpha_n$ ($\alpha_p$) is $1.67 \times 10^{11}$ cm$^{-2}$V$^{-1}$ ($5.7 \times 10^{9}$ cm$^{-2}$ V$^{-1}$). For both devices at $T = 1.8$ K, the hole mobility exceeds the electron mobility (see Fig. 1 in the Supplemental Material).

Figure 2(a) shows the Hall conductivity $\sigma_{yx}(V_g)$ for Device 1 in various magnetic fields ranging from 1.5 – 4.0 T. According to the QHE theory $\sigma_{yx} = \nu e^2/h$, where the filling factor ν determines the number of the filled LLs. Thus, transitions between quantum plateaus correspond to crossings of the Fermi level with the LL centers. Figure 2(b) presents the $\sigma_{yx}$ mapping as a function of the gate voltage $V_g$ and the magnetic field $B$ at $T = 1.8$ K. The QHE plateaus at ν = 1 and ν = –1 are marked by yellow and red areas and correspond to electron and hole transport, respectively. One can see that the QHE plateaus for holes (ν = –1) are much wider than those for electrons (ν = 1).

We present experimental studies of magnetoresistances in such HgTe quantum wells versus temperature and magnetic field, which allow determining the critical magnetic field $B_c$ above which the topological gap (and parasitic edge conduction channels) disappears. We show that for QWs for which $B_c$ becomes very low or even vanishes, one observes remarkably wide QHE plateaus at filling factor ν = – 1 (holes) using only commercial Neodymium permanent magnets providing $B$ = 0.82 T, and liquid $^4$He system operating at a few Kelvins.

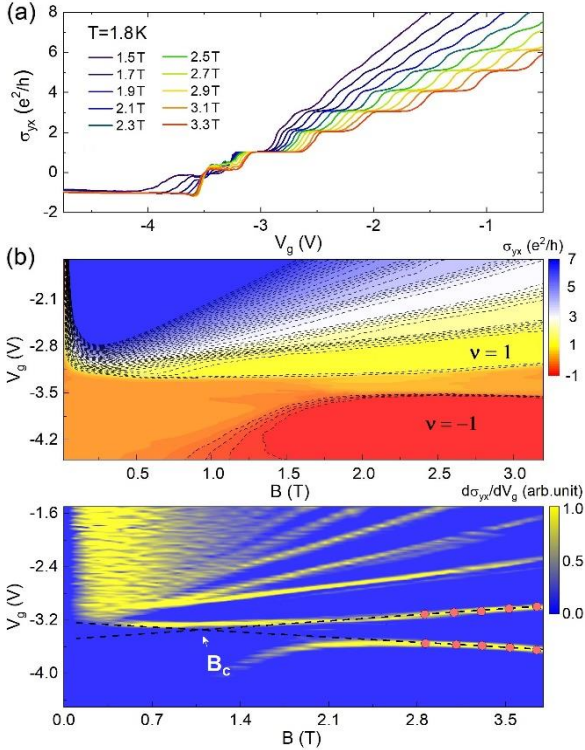

derivative $d\sigma_{yx}/dV_g$, where $\sigma_{yx} = \rho_{xy}/(\rho_{xx}^2 + \rho_{xy}^2)$. Symbols correspond to the onset of the regions with $\sigma_{yx} = \pm e^2/h$. The dashed lines are the guides for the eyes for the expected position of the zero-mode LLs.

It is worth noting that the transition from inverted to normal band ordering occurs with the increase of the magnetic field. The derivative of the Hall conductivity, shown in Fig. 2(c), exhibits sharp peaks corresponding to the interception of LLs by the Fermi level. The two peaks arising from the transitions correspond to the two zero-mode LLs, shown in Fig. 1. Mapping $d\sigma_{yx}/dV_g$ versus $V_g$ and $B$ allows obtaining the graph corresponding to LL fan chart. The transition from the inverted to normal band ordering for device D1 occurs around $B_c \simeq 1.2$ T [see Fig. 2(c)]. This means that the quantum spin Hall effect is eliminated for magnetic fields higher than $B_c \simeq 1.2$ T and replaced by the ordinary integer QHE [14,20].

For ordinary semiconductor systems, like GaAs/AlGaAs heterostructures, either the fractional plateaus or vanishing Hall conductivity ($\sigma_{xy} = 0$), are expected in high fields when the filling factor diminishes well below 1. For HgTe QWs, this QHE regime is observed in magnetic fields exceeding a critical value $B > B_c$, and for gate voltages in the vicinity of the charge neutrality point. At $B = B_c$, the QHE behavior is analogous to the case of massless Dirac fermions, with a direct transition between two states with $\sigma_{yx} = \pm e^2/h$.

We present now the magnetotransport properties of device D2. In Figs. 3(a,b) we depict the Hall resistivity $\rho_{xy}(B)$ as a function of electron and hole densities controlled by the gate voltage. As seen, the QHE plateaus ν = ±1 appear at a lower magnetic field for holes than for electrons.

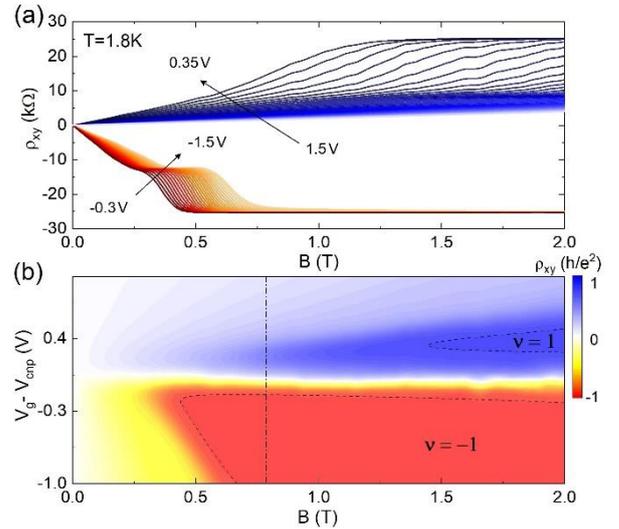

**Fig. 3.** (a) Device D2. Hall resistivity for electrons and holes at various gate voltages $V_g - V_0$. (b) Map of the Hall resistivity as a function of the magnetic field for different gate voltages with respect to the charge neutrality point. QHE plateaus at ν = ±1 correspond to blue and red areas for electrons and holes, respectively. Vertical dashed-dot line corresponds B =0.82 T.

**Fig. 2.** (a) Device D1. Dependence of the Hall conductivity on the gate voltage in various magnetic fields at $T$ = 1.8 K. (b) Map of the Hall conductivity (with contours and color scale) as a function of the magnetic field and gate voltage. The dashed lines are regularly spaced lines of the Hall conductivity. QHE plateaus at ν = ±1 correspond to yellow and red areas for electrons and holes, respectively. (c) Experimentally obtained Landau levels as a function of the gate voltage $V_g$. Color map shows the absolute values of the

Overall, there are two prerequisites for finding a precise quantization: $\mu B \gg 1$ and $\Delta > 100\, k_B T$, i.e., large values of both carrier mobility μ and gap between neighboring Landau levels Δ. Compared to other materials [4], HgTe/HgCdTe QWs exhibit sizable carrier mobility (~$10^5$ cm²/Vs) and large gap ($\Delta \simeq 10$ meV). Accordingly, in device D2 quantized Hall conductance of holes appears already in $B = 0.5$ T at $T = 2.0$ K. Strikingly, in both devices, the quantized Hall plateaus are much wider for holes than for electrons. This may be explained by the existence of the valence band side maxima serving as charge reservoir specific to HgTe quantum wells.

The features of the QHE in HgTe QWs are strongly correlated to the behavior of the zero-mode LLs. Therefore, another objective of the present work is to investigate the properties of these zero-mode LLs *versus* temperature.

The color maps shown in Fig. 4 evidence the transition from the inverted band ordering to the normal band sequence with increasing temperature. This is confirmed by theoretical computations that show that with increasing temperature, the topological energy gap between the electron-like subband E1 and the heavy hole-like subband H1 decreases so that above critical temperature $T_c$ a normal band ordering appears. Experimentally, according to results displayed in Fig. 4(b), $T_c = 10$ K for device D2. A further increase of temperature results in a gap opening, see Fig. 4(c). It is worth noting that analogous changes in band structure can be caused by hydrostatic pressure [19]. Finally, in agreement with Ref. 26, our results indicate that the carrier density does not change with temperature (see Fig. 2 in the Supplemental Material).

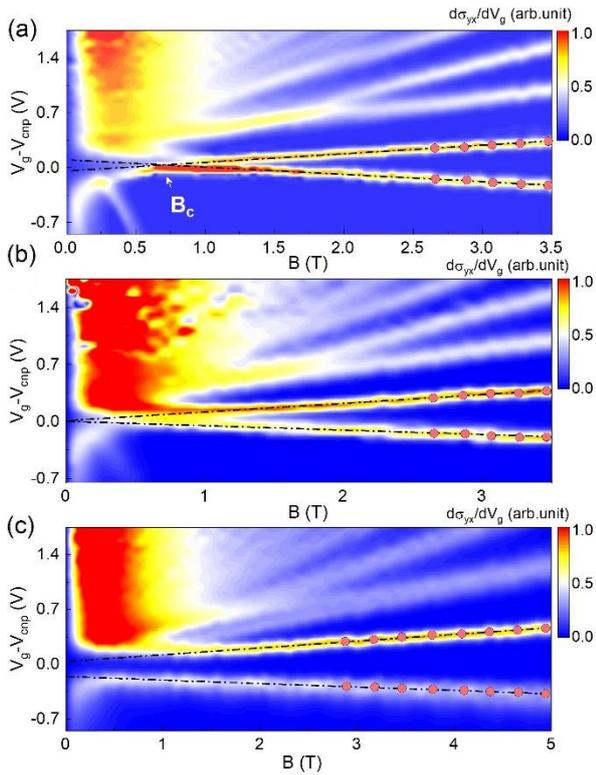

**Fig. 4.** Device D2. Experimentally obtained Landau level fan charts as a function of the gate voltage $V_g$ at various temperatures: (a) $T = 1.8$ K; (b) $T = 10$ K and (c) $T = 20$ K. Color maps depict the absolute values of the derivative $d\sigma_{xy}/dV_g$. Symbols correspond to the onset of the regions with $\sigma_{yx} = \pm e^2/h$. The dashed-dot lines show the zero-mode LLs.

As can be seen on Fig. 3b, the plateau $\nu = -1$ starts from fields about 0.5 T. Such fields are available using permanent Nd magnets. In order to construct a prototype device for resistance standard we have prepared a specific sample holder equipped with permanent magnets as shown in Figs. 5(a,b).

The results shown in Fig. 5(c) clearly demonstrate the appearance of QHE in device D2 in the magnetic field produced by permanent magnets. The Hall resistance as a function of gate voltage is measured in the temperature range $2.0 – 5.0$ K. Quantum Hall plateau $\nu = -1$ is observed up to 15 K, while the Fermi level remains pined in the valence band. Under our experimental conditions, we can only roughly estimate the accuracy of $\rho_{xy}$ quantization as $\pm 0.01 h/e^2$ at $T = 1.6$ K and a minimum value of $\rho_{xx}$ is below $0.001 h/e^2$ at $T = 1.6$ K and for $I = 100$ nA.

The robustness of the QHE states in the hole transport regime is presumably related to the Fermi level pinning by the side maxima in the valence band, playing the role of a charge reservoir [20,27,28].

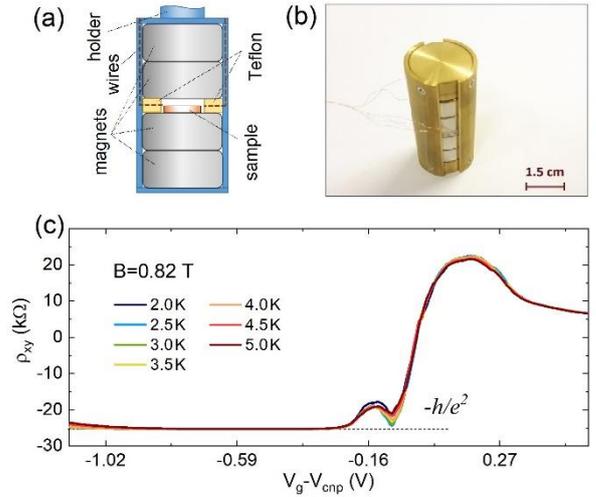

**Fig. 5.** (a) Schematic illustration and (b) photography of the sample holder with four permanent Nd magnets. (c) Hall resistance as a function of the gate voltage in the magnetic field of 0.82 T generated by the permanent magnets for different temperatures: 2.0 K (navy blue), 2.5 K (blue), 3.0 K (green), 3.5 K (yellow), 4.0 K (orange) 4.5 K (red) and 5.0 K (brown).

## IV. SUMMARY AND FUTURE DEVELOPEMENTS

To summarize, gated HgTe quantum wells of width around 7 nm exhibit graphene like band structure with low effective mass ("masless Fermions") and high hole mobilities. These properties grant well defined LLs with large energy gaps between them even at relatively low magnetic fields. These peculiar properties have allowed us to observe a wide $\nu = -1$ QHE plateaus using Nd permanent magnets. We also performed transport measurements of HgTe QWs at various temperatures, revealing a phase transition at which the two-dimensional Dirac massless fermions emerge at $T = 10$ K. Our results demonstrate that these HgTe quantum structures are potential candidates for quantum Hall resistance standards operating in relaxed cryomagnetic conditions

There are a few challenges in further development of HgTe QW resistance standards. One of them is the reduction of the contact resistance, which induces local heating and precludes the use of high currents needed to reach a better metrological precision [29,30].

We suggest increasing carrier density under the contacts by using an additional gate in the contact proximity.

Also, one can consider a conductive substrate serving as a HgTe QWs bottom gate [31]. Alternatively, one could also consider double HgTe QWs, where one of the QWs plays the role of the gate controlling the carrier concentration. Moreover, metals like Sn, In, or Au/Mo are required to form low-resistive ohmic contacts [32,33], precluding Schottky barrier formation.

The band structure properties of HgTe QWs can be modified to obtain metrological QHRSs. It has been demonstrated theoretically that the positions of extrema in the valence band dispersion and the magnitude of LL gap can be controlled by axial strain [20,34,35]. Additionally, the band structure properties can be manipulated by adding Mn and Cd into the QWs barriers [28,34]. These additional degrees of freedom may increase the width and the robustness of the QHE plateaus, allowing further improvement of HgTe QWs based metrological devices operating with relaxed cryomagnetic conditions.

## ACKNOWLEDGEMENTS

Research in CENTERA Laboratories and in the International Research Centre MagTop was supported in the framework of the International Research Agendas Program was supported of the Foundation for Polish Sciences co-financed by the European Union under the European Regional Development Fund (MAB/2018/9 and MAB/2017/1), by the TEAM project POIR.04.04.00-00-3D76/16 (TEAM/2016-3/25), National Science Centre Poland (grant No. UMO-2017/25/N/ST3/00408) and by CNRS through IRP "TeraMIR".

## REFERENCES

[1] K. v. Klitzing, G. Dorda, M. Pepper, New Method for High-Accuracy Determination of the Fine-Structure Constant Based on Quantized Hall Resistance, Phys. Rev. Lett. **45,** 494 (1980).
[2] D. J. Thouless, M. Kohmoto, M. P. Nightingale, M. den Nijs, Quantized Hall conductance in a two-dimensional periodic potential. Phys. Rev. Lett. **49,** 405–408 (1982).
[3] B. Jeckelmann and B. Jeanneret, The quantum Hall effect as an electrical resistance standard, Rep. Prog. Phys. **64,** 1603–1655 (2001).
[4] A.F. Rigosi and R. E. Elmquist, The quantum Hall effect in the era of the new SI, Semicond. Sci. Techn. **34,** 093004 (2019) .
[5] A. Hartland, K. Jones, J. M. Williams, B. L. Gallagher, and T. Galloway, Direct comparison of the quantized Hall resistance in gallium arsenide and silicon, Phys. Rev. Lett. **66,** 969 (1991).
[6] B. Jeckelmann, B. Jeanneret, and D. Inglis, High-precision measurements of the quantized Hall resistance:Experimental conditions for universality, Phys. Rev. B **55,** 13124 (1997).
[7] F. Schopfer and W. Poirier, Quantum resistance standard accuracy close to the zero-dissipation state, J. Appl. Phys. **114,** 064508 (2013) .
[8] A, Tzalenchuk, S. Lara-Avila, A. Kalaboukhov, S. Paolillo, M. Syväjärvi, R. Yakimova, O. Kazakova, T. J. B. M. Janssen, V. Fal'ko, and Sergey Kubatkin., Towards a quantum resistance standard based on epitaxial graphene. Nat. Nano. **5,** 186 - 189 (2010).

[9] R. Ribeiro-Palau, F. Lafont, J. Brun-Picard, D. Kazazis, A. Michon, F. Cheynis, O. Couturaud, C. Consejo, B. Jouault, W. Poirier, F. Schopfer, Quantum Hall resistance standard in graphene devices under relaxed experimental conditions, Nat. Nano. **10,** 965–971 (2015).
[10] F. Lafont, R. Ribeiro-Palau, D. Kazazis, A. Michon, O. Couturaud, C. Consejo, T. Chassagne, M. Zielinski, M. Portail, B. Jouault, F. Schopfer & W. Poirier, Quantum Hall resistance standards from graphene grown by chemical vapour deposition on silicon carbide, Nat. Commun. **6,** 6806 (2015).
[11] K. S. Novoselov, E. McCann, S. V. Morozov, V. I. Fal'ko, M. I. Katsnelson, U. Zeitler, D. Jiang, F. Schedin & A. K. Geim, Unconventional quantum Hall effect and Berry's phase of $2\pi$ in bilayer graphene, Nature Physics **2,** 177–180 (2006).
[12] L.G. Gerchikov and A.V. Subashiev, Interface States in Subband Structure of Semiconductor Quantum Wells, Phys. Stat. Sol. B **160,** 443 (1990).
[13] B. Büttner, C. X. Liu et al., Single valley Dirac fermions in zero-gap HgTe quantum wells, Nat. Phys.**7,** 418–422 (2011).
[14] M. Konig, S. Wiedmann, C. Brune, A. Roth, H. Buhmann,L. W. Molenkamp, X. L. Qi,and S. C. Zhang, Quantum Spin Hall Insulator State in HgTe Quantum Wells , Science **318,** 766 (2007).
[15] P. Olbrich, C. Zoth, P. Vierling, K.-M. Dantscher, G. V. Budkin, S. A. Tarasenko, V. V. Belkov, D. A. Kozlov, Z. D. Kvon, N. N. Mikhailov, S. A. Dvoretsky, and S. D. Ganichev, Giant photocurrents in a Dirac fermion system at cyclotron resonance, Phys. Rev. B **87,** 235439 (2013)
[16] D. A. Kozlov, Z. D. Kvon, N. N. Mikhailov, S. A. Dvoretskii, Quantum Hall effect in a system of gapless Dirac fermions in HgTe quantum wells, JETP Lett.**100,** 724 (2015).
[17] D. A. Kozlov, Z. D. Kvon, N. N. Mikhailov, S. A. Dvoretskii, S. Weishäupl, Y. Krupko, J. C. Portal, Quantum Hall effect in HgTe quantum wells at nitrogen temperatures, Appl. Phys. Lett. **105,** 132102 (2014).
[18] T. Khouri, M. Bendias, P. Leubner, C. Brune, H. Buhmann, L. W. Molenkamp, U. Zeitler, N. E. Hussey, and S. Wiedmann, High-temperature quantum Hall effect in finite gapped HgTe quantum wells, Phys. Rev. B **93,** 125308 (2016).
[19] S. S. Krishtopenko, I. Yahniuk, D. B. But, V. I. Gavrilenko, W. Knap, and F. Teppe, Pressure- and temperature-driven phase transitions in HgTe quantum wells, Phys. Rev. B **94,** 245402 (2016).
[20] I. Yahniuk, S.S. Krishtopenko, G. Grabecki, B. Jouault, C. Consejo, W. Desrat, M. Majewicz, A.M. Kadykov, K.E. Spirin, V.I. Gavrilenko, N.N. Mikhailov.. Magneto-transport in inverted HgTe quantum wells, npj Quant. Mater. **4,** 13 (2019).
[21] F. Parmentier, T. Cazimajou, Y. Sekine, H. Hibino, H. Irie, D. Glattli, N. Kumada, and P. Roulleau, Quantum Hall effect in epitaxial graphene with permanent magnets, Sci. Rep. **6,** 38393 (2016).
[22] M. Götz, K. M. Fijalkowski, E. Pesel, M. Hartl, S. Schreyeck, M. Winnerlein, S. Grauer, H. Scherer, K. Brunner, C. Gould, F. J. Ahlers, and L. W. Molenkamp, Precision measurement of the quantized anomalous Hall resistance at zero magnetic field, Appl. Phys. Lett. **112,** 072102 (2018).
[23] E. J. Fox, I. T. Rosen, Yanfei Yang, G. R. Jones, R. E. Elmquist, Xufeng Kou, Lei Pan, Kang L. Wang, D. Goldhaber-Gordon, Part-per-million quantization and current-induced breakdown of the quantum anomalous Hall effect, Phys. Rev. B **98,** 075145 (2018).
[24] S. Dvoretsky, N. Mikhailov, Y. Sidorov, V. Shvets, S. Danilov, B. Wittman, and S. Ganichev, Growth of HgTe

quantum wells for IR to THz detectors, J. Electron. Mater. **39**, 918 (2010).

[25] M. Majewicz, D. Śnieżek, T. Wojciechowski, E. Baran, P. Nowicki, T. Wojtowicz, J. Wróbel, Low temperature processing of nanostructures based on II-VI semiconductors quantum wells, Acta. Phys. Pol. A **126**, i 1174-1176 (2014).

[26] A. M. Kadykov, S. S. Krishtopenko, B. Jouault, W. Desrat, W. Knap, S. Ruffenach, C. Consejo, J. Torres, S. V. Morozov, N. N. Mikhailov, S. A. Dvoretskii, and F. Teppe, Temperature-induced topological phase transition in HgTe quantum wells, Phys. Rev. Lett. **120**, 086401 (2018).

[27] S. Mantion, C. Avogadri, S. Krishtopenko, S. Gebert, S. Ruffenach, C. Consejo, and B. Jouault, Quantum Hall states in inverted HgTe quantum wells probed by transconductance fluctuations, Phys. Rev. B **102**, 075302 (2020).

[28] S. Saquib, W. Beugeling, J. Böttcher, P. Shekhar, A. Budewitz, P. Leubner, L. Lunczer, E.M. Hankiewicz, H. Buhmann, and L.W. Molenkamp, Emergent quantum Hall effects below 50 mT in a two-dimensional topological insulator, Sci. Adv. **6**, eaba4625 (2020).

[29] J. Weis and K. Von Klitzing, Metrology and microscopic picture of the integer quantum Hall effect, Phil. Trans. R. Soc. A **369**, 3954–3974 (2011).

[30] R. J. Haug, Edge-state transport and its experimental consequences in high magnetic fields, Semicond. Sci. Technol. **8**, 131-153 (1993).

[31] M. Baenninger, M. König, A. G. F. Garcia, M. Mühlbauer, C. Ames, P. Leubner, C. Brüne, H. Buhmann, L. W. Molenkamp, and D. Goldhaber-Gordon, Fabrication of samples for scanning probe experiments on quantum spin Hall effect in HgTe quantum wells, J. Appl. Phys. **112**, 103713 (2012).

[32] P. W. Leech, The specific contact resistance of Ohmic contacts to HgTe/Hg$_{1-x}$Cd$_x$Te heterostructures, J. Appl. Phys. **68**, 907 (1990).

[33] Dan Liu, Chun Lin, Songmin Zhou, and Xiaoning Hu, Ohmic contact of Au/Mo on Hg$_{1-x}$Cd$_x$Te, J. Electron. Mater. **45**, 6 (2016).

[34] S. Parijat, T. Kubis, Y. Tan, M. Povolotskyi, and G. Klimeck, Design principles for HgTe based topological insulator devices, J. Appl. Phys. **114**, 043702 (2013).

[35] P. Leubner, L. Lunczer, C. Brüne, H. Buhmann, L.W. Molenkamp. Strain Engineering of the Band Gap of HgTe Quantum Wells Using Superlattice Virtual Substrates, Phys. Rev. Lett. **117**, 086403 (2016)

# Supplemental Materials

**HgTe quantum wells for QHE metrology with relaxed cryomagnetic conditions: permanent magnets and liquid $^4$He temperatures**

I. Yahniuk,[1] A. Kazakov,[2] B. Jouault,[3] S. S. Krishtopenko,[1,3] S. Kret,[4] G. Grabecki,[4] G. Cywiński,[1] N. N. Mikhailov,[5] S. A. Dvoretskii,[5] J. Przybytek,[1] V. I. Gavrilenko,[6] F. Teppe,[3] T. Dietl,[2,7] and W. Knap[1, 3, *]

[1]CENTERA Laboratories, Institute of High-Pressure Physics, Polish Academy of Sciences, Warsaw, PL-01142 Poland

[2]International Research Centre MagTop, Institute of Physics, Polish Academy of Sciences, al. Lotników 32/46, PL-02668, Warsaw, Poland.

[3]Laboratoire Charles Coulomb, Université de Montpellier, Centre National de la Recherche Scientifique, F-34095 Montpellier, France.

[4]Institute of Physics, Polish Academy of Sciences, al. Lotników 32/46, PL-02668 Warsaw, Poland.

[5]Institute of Semiconductor Physics, Siberian Branch, Russian Academy of Sciences, pr. Akademika Lavrent'eva 13, Novosibirsk, 630090 Russia.

[6]Institute for Physics of Microstructures, Russian Academy of Sciences, 603950 N. Novgorod, Russia.

[7]WPI-Advanced Institute for Materials Research, Tohoku University, Sendai 980-8577, Japan

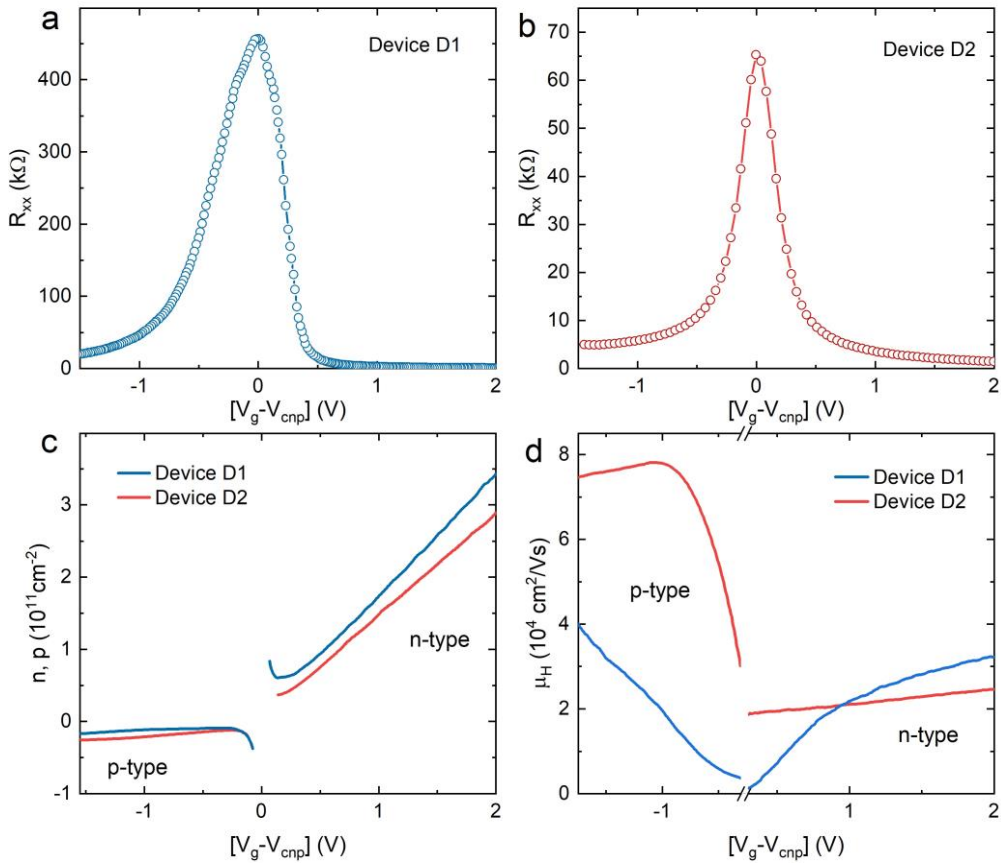

Fig. 1. The longitudinal resistance in device D1 (a) and device D2 (b) as a function of the gate voltage at T=1.8 K. (c) Hall carrier concentration and (d) Hall mobility in D1 and D2 for electrons and holes versus gate voltage at T=1.8 K.

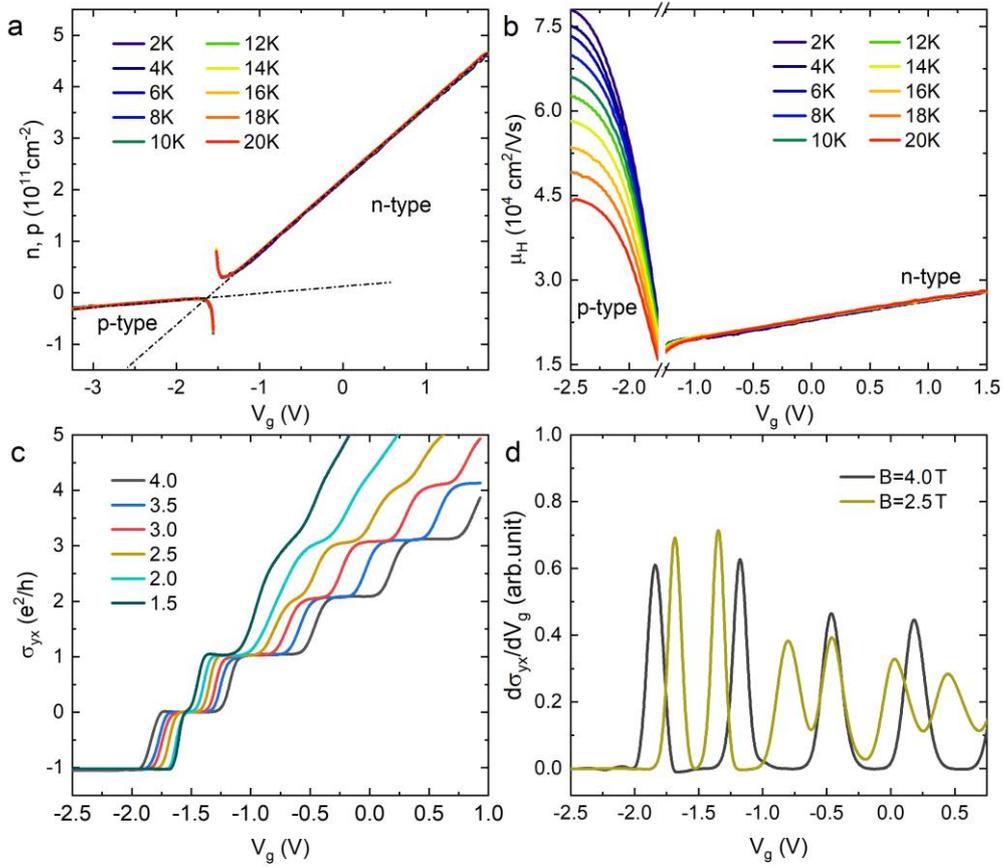

Fig. 2 (a) Hall carrier concentration and (b) Hall mobility for electrons and holes as a function of gate voltage at temperature range from 2 K to 20 K. (c) Dependence of the Hall conductivity on gate voltage presented for different magnetic fields: (dark cyan) 1.5 T; (cyan) 2.0 T;(dark yellow) 2.5 T; (red) 3:0 T ; (blue) 3.5 T and (dark gray) 4.0 T. (d) Derivative of the transverse conductivity for (dark goldenrod curve) 2.5 T and (gray curve) 4.0 T from the gate voltage.